\documentstyle[12pt]{article}
\makeatletter
\newcommand{\singlespacing}{\let\CS=\@currsize\renewcommand{\baselinestretch}
{1.0}\tiny\CS}
\newcommand{\doublespacing}{\let\CS=\@currsize\renewcommand{\baselinestretch}
{1.5}\tiny\CS}
\newcommand{\ed}{\end{document}}
\newcommand{\bc}{\begin{center}}
\newcommand{\ec}{\end{center}}
\newcommand{\bfr}{\begin{flushright}}
\newcommand{\efr}{\end{flushright}}

\newcommand{\beq}{\begin{equation}}
\newcommand{\eeq}{\end{equation}}

\newcommand{\ben}{\begin{enumerate}}
\newcommand{\een}{\end{enumerate}}
\newcommand{\bit}{\begin{itemize}}
\newcommand{\eit}{\end{itemize}}

\newcommand{\ba}{\begin{array}}
\newcommand{\ea}{\end{array}}
\newcommand{\beqa}{\begin{eqnarray}}
\newcommand{\eeqa}{\end{eqnarray}}
\newcommand{\beqas}{\begin{eqnarray*}}
\newcommand{\eeqas}{\end{eqnarray*}}
\newcommand{\bfg}{\begin{figure}}
\newcommand{\efg}{\end{figure}}
\newcommand{\bds}{\begin{displaymath}}
\newcommand{\eds}{\end{displaymath}}
\newcommand{\btb}{\begin{tabbing}}
\newcommand{\etb}{\end{tabbing}}

\newcommand{\pad}{\partial}

\newcommand{\g}{\gamma}

\newcommand{\eps}{\epsilon}

\newcommand{\s}{\sigma}

\newcommand{\pr}{\prime}

\newcommand{\m}{\mu}
\newcommand{\n}{\nu}

\newcommand{\p}{\pi}

\newcommand{\f}{\frac}

\begin{document}
\begin{center}{\large{\bf Pairing in High Temperature Superconductors
and Berry Phase}}
\end{center}
\begin{center}
{\bf D.Pal}\
\footnote{e-mail : res9719@www.isical.ac.in}~and~{\bf B.Basu}\footnote
{ e-mail : banasri@www.isical.ac.in}\\
{\bf Physics and Applied Mathematics Unit}\\
{\bf Indian Statistical Institute}\\
{\bf Calcutta-700035}
\end{center}
\date{}


\vspace*{1cm}


\centerline{\bf Abstract}
The topological approach to the understanding of pairing mechanism
in high $T_c$ superconductors analyses
the relevance of the Berry phase factor in this context. This also gives the
evidence for the pairing mechanism to be of magnetic origin.

\vspace*{1cm}
\thispagestyle{empty}

The origin and nature of the high $T_c$ superconductivity of the
$La_{2-x}M_xCuO_{4-\delta}$ and $RBa_2Cu_3O_{7-\delta}$ compounds
(M=Ba, Sr, Ca, Na \ldots..and R=Y, La, Nd,\ldots), as well as the other layered
copper oxide compounds, is not presently understood and constitutes a
formidable challenge to experimentalists and theorists alike.
It is known that there is proximity
of antiferromagnetism and superconductivity as the concentration of holes in
the conducting $CuO_2$ planes is varied. This suggests the primary evidence
for the pairing mechanism to be of magnetic origin.
 In this note we shall
study the pairing mechanism through our understanding of high $T_c$
superconductivity in terms of Berry phase as formulated in an earlier
paper [1].
We shall show that a magnetic type of gauge interction is responsible for the
formation of the pair leading to superconductivity.

In a recent paper [1] we have analysed the equivalence of Resonating Valence
Bond (RVB) state with fractional quantum Hall fluid with filling factor
$\n=1/2$ in terms of Berry phase which is associated with the
chiral anomaly in 3+1 dimensions. It is noted that the three dimensional
spinons and holons are characterized by the non-Abelian Berry phase and these
reduce to $1/2$ fractional statistics when the motion is confined to
equatorial planes. We have shown that the topological mechanism of
superconductivity is analogous to the topological aspects of fractional
quantum Hall effect with $\n=1/2$. Our result corroborates with the idea of
Laughlin [2a,2b]. In our framework it is argued that
a frustrated spin system on a lattice is
characterized by the chirality which is associated with the Berry phase factor
$\m$ where the phase is given by $e^{i 2 \p \m}$. We know that the ground
state of an
antiferromagnet is characterized by two operators, namely density of energy
\beq
\eps_{ij} = ( \f{1}{4} + \vec{S_i}. \vec{S_j})
\eeq
and chirality
\beq
W(C) = Tr \prod_{i \in C} ( \f{1}{2} + \vec{\s} . \vec{S_i})
\eeq
where $\s$ are Pauli matrices and $C$ is a lattice contour. Weigman [3] has
related these operators with the amplitude and phase $\Delta_{ij}$ of
Anderson's RVB through
\beq
\eps_{ij} = {| \Delta_{ij} |}^2
\eeq
\beq
W(C) = \prod_C \Delta_{ij}
\eeq
This suggests that $\Delta_{ij}$ is a gauge field. The topological order
parameter $W(C)$ acquires the form of a lattice Wilson loop
\beq
W(C) = e^{i \phi (c)}
\eeq
which is associated with the flux of the RVB field
\beq
e^{i \phi (c)} = \prod_C e^{i A_{ij}}
\eeq
$A_{ij}$ is a phase of $\Delta_{ij}$ representing a magnetic flux which
penetrates through a surface enclosed by the contour $C$. This phase is
essentially the Berry phase related to chiral anomaly when we describe the
system in three dimensions through the relation
\beq
W(C) = e^{i 2 \pi  \m}
\eeq
where $\m$ appears to be a monopole strength.
In view of this, when a two dimensional frustrated spin system on a lattice
resides on
the surface of a three dimensional sphere of a large radius in a radial
(monopole) magnetic field, we can associate the chirality with the Berry
phase.

It is wellknown that the system of correlated electrons on a lattice is
governed
by the Hubbard model which in the strong coupling limit and at half filling
can be mapped onto an antiferromagnetic Heisenberg model with nearest
neighbour interaction which is represented by the Hamiltonian
\beq
H = J \sum ( S^x_i S^x_j + S^y_i S^y_j + S^z_i S^z_j)
\eeq
with $J>0$. As we have mentioned earlier, a frustrated spin system leading to
RVB state is characterised by the chirality associated with it. In view of
this we can consider an anisotropic Heisenberg Hamiltonian with nearest
neighbour interaction which may be represented as
\beq
H = J \sum ( S^x_i S^x_j + S^y_i S^y_j + \Delta S^z_i S^z_j)
\eeq
where $J > 0$ and $\Delta \ge 0$ and the anisotropy parameter $\Delta$ is
given by $\Delta = \f{2\m + 1}{2}$ [4], $\m$ being the Berry phase factor
and $\m$ can take the values $\m = 0, \pm 1/2, \pm 1, \pm 3/2
........$ .
We note  that for $\Delta = 1$ corresponding to $\m = 1/2$ one has the
isotropic antiferrromagnetic Heisenberg model. When $\Delta = 0 (\m = - 1/2)$
we have
the $XX$ model. It may be mentioned here that the relationship of the
anisotropy parameter $\Delta$ with the Berry phase facter $\m$ has been
formulated from an analysis of the relationship between the conformal field
theory in $1+1$ dimension, Chern-Simons theory in $2+1$ dimension and chiral
anomaly in $3+1$ dimension [4].

Now for a two dimensional frustrated spin system on a lattice on the surface
of a three
dimensional sphere in a radial magnetic field the chirality demands that $\m$
is non-zero and should be given by $|\m| = \f{1}{2}$ in the ground state.
However for $\m = 1/2$, we get the isotropic antiferromagnetic Heisenberg
Hamiltonian. Now let us consider the ground state of antiferromagnetic
Heisenberg model on a lattice which allows frustration to occur
giving rise to resonating valence bond (RVB) states corresponding to spin
singlets
where two nearest neighbour bonds are allowed to resonate among themselves.
This situation occurs with $\m=-1/2$
 suggesting $\Delta = \f{2\m+1}{2}=0$ in the Hamiltonian (9). But with
$\Delta =0$, the Hamiltonian effectively represents $XX$ model corresponding
to a bosonic system represented by singlets of spin pairs. This eventually
leads to a resonating valence bond state giving rise to a nondegenerate quantum
liquid. We have argued [1] that these spin singlet states forming the
quantum liquid are equivalent to fractional quantum Hall (FQH) liquid with
filling factor $\n = 1/2$.
Indeed, in earlier papers [5,6] we have pointed out that in QHE the
external magnetic field causes the chiral symmetry breaking of the fermions
(Hall particles) and an anomaly is realized in asssociation with the
quantization of Hall conductivity. This helps us to study the behaviour of a
quantum Hall fluid from the viewpoint of the Berry phase which is linked with
chiral anomaly when we consider a 2D electron gas of N-particles on the
surface of a three dimensional sphere in a radial (monopole)
strong magnetic
field.  For the FQH state with $\n = 1/2$ [5], the
Dirac quantization condition $e \m = 1/2$ suggests that $\m=1$. Then in the
angular momentum relation for the motion of a charged particle in the field of
a magnetic monopole
\beq
\vec{J} = \vec{r} \times \vec{p} - \m \vec{r}
\eeq
we note that for $\m = 1$ ( or an integer) we can use a transformation which
effectively suggests that we can have a dynamical relation of the form
\beq
\vec{J} = \vec{r} \times \vec{p} - \m \vec{r} = \vec{r^\pr} \times \vec{p^\pr}
\eeq
This indicates that the Berry phase which is associated with $\m$ may be
unitarily removed to the dynamical phase. This implies that the average
magnetic field may be taken to be vanishing in these states. However, the
effect of the Berry phase may be observed when the state is split into a pair
of electrons each with a constraint of representing the state with $\m = \pm
1/2$. These pairs will give rise to the $SU(2)$ symmetry as we can consider
the state of these two electrons as a $SU(2)$ doublet. This doublet of Hall
particles
 for $\n = 1/2$ FQH fluid may be considered to be equivalent to RVB singlets.

It is known that the RVB states are characterized by neutral spin $\f{1}{2}$
excitations called spinons and charged spinless excitations called holons.
To study these excitations in our Berry phase approach, we have to consider a
 system  such that a two dimensional frustrated spin
system lies on the surface of a three dimensional sphere in a radial
(monopole) magnetic field which gives rise to the chirality associated with
the monopole. As discussed above a RVB state may be characterised by
associating a magnetic flux at the background corresponding to $\m=-1/2$ in the
Berry phase factor. Now if we consider a  single spin down electron (with a
$\m=-1/2 $ value) at a site $j$  surrounded by an otherwise featureless
spin liquid representing a RVB state the associated magnetic flux at
the site $j$ may be characterised by the monopole strength $\m=-1/2$ which
represents the Berry phase factor for the isolated spin down electron at the
site $j$. Indeed, a spin polarised electron may be represented by a two
component spinor such that spin down and up states correspond to particle and
antiparticle and from the relation $e \m=1/2$, we may associated spin down and
up states with $\m=-1/2$ and $+1/2$. Now the magnetic flux associated with the
spin down electron at the site $j$ when combined with the magnetic flux at the
background due to the frustration of the spin system, we find the effective
Berry phase factor is given by $\m =-1$. This means that when a spinon moves
in a closed path represented by a plaquette it will acquire the phase factor
$|\m|=1$  for the Berry phase given by $e^{i2 \pi \m}$.
 The units of magnetic flux
associated with the Berry phase factor $\m$ may be represented through a phase
\beq
e^{i 2 \p \m} = \prod_C e^{i A_{ij}}
\eeq
where $A_{ij}$ is a phase representing a magnetic flux which penetrates through
a surface enclosed by the contour $C$.
 As pointed out earlier, the Berry phase factor $|\m|=1$ effectively
gives rise to a FQH state with $\n =\f{1}{2}$.  Thus
the neutral spin $\f{1}{2}$ excitation, the spinon is characterized by
$|\m|=1$  which may be split into two
parts, with one spin $\f{1}{2}$ excitation in the bulk and the other part is
due to the {\it orbital spin} which is in the background characterized by the
chirality of the frustrated spin system.
The relation $e \m = \f{1}{2}$ suggests that the spinon
effectively represents a particle having $1/2$ fractional statistics in a 2D
system which is analogous to the idea of Laughlin [2].

 For the spin-charge recombination
we have formulated a similar
formalism of Weng, Sheng and Ting [7]. They have considered the spin-charge
separation saddle-point by the effective Hamiltonian
\beq
H_{eff} = H_s + H_h
\eeq
where $H_s$ and $H_h$ are the spinon and holon Hamiltonian respectively,
defined by
\beq
H_s = -J_s \sum_{<ij> \s} (e^{i \s A^h_{ij}}) b^\dagger_{i \s} b_{j \s} + h.c
\eeq
\beq
H_h = -t_h \sum_{<ij>} e^{i( - \phi^0_{ij} + A^s_{ij})} h^\dagger_i h_j +
h.c
\eeq
where $b_{i \s}$ and $h_i$ are spinon and holon annihilation operators
respectively. Here $\phi^0_{ij}$ represents flux quanta threading through
each plaquette. The topological phases $A^h_{ij}$ and $A^s_{ij}$ are such that
$A^h_{ij}$ describes flux quanta attached to holons which are seen only by
spinons in $H_s$ and $A^s_{ij}$ describes flux quanta bound to spinons which
can only be seen by holons. Thus $A^h_{ij}$ represents the doping effect into
spin degrees of freedom where $A^s_{ij}$ plays the role of a scattering source
in holon transport.

In our framework, we can write the effective Hamiltonian for the spinon as
\beq
H_s = -J_s \sum_{<ij> \s} (e^{i \s A_{ij}}) b^\dagger_{i \s} b_{j \s} + h.c
\eeq
where $b_{i \s}$ is the spinon annihilation operator and $A_{ij}$ represents
the magnetic flux penetrating through a surface enclosed by a contour $C$ and
is given by eqn.(12).
Similarly we can write a Hamiltonian for holon
\beq
H_h = -t_h \sum_{<ij>} e^{i( - \phi^0_{ij} + A_{ij})} h^\dagger_i h_j +
h.c
 \eeq
where $b_{i \s}$ and $h_i$ are spinon and holon annihilation operators
respectively. Here $\phi^0_{ij}$ represents flux quanta threading through
each plaquette.
The interaction between spinons and holons are then mediated through the
gauge field $A_{ij}$.

When a hole is introduced, the spinon will interact with the hole through the
propagation of the magnetic flux and this coupling will lead to the creation
of the holon having magnetic flux corresponding to $|\m_{eff}|=1$. Eventually,
 the residual spinon will be
devoid of any magnetic flux corresponding to $\m_{eff}=0$. This is realized
when the unit of magnetic flux $\m = - \f{1}{2}$ associated with the single
down spin in the RVB liquid will form a pair with another up spin having $\m =
+ \f{1}{2}$ associated with the hole. Thus the holon characterised by
  $|\m_{eff}| = 1$ represent a spinless charged particle having $\f{1}{2}$
fractional statistics in a 2D system.  Indeed, the isolated
spin in the RVB liquid will be combined with the spin associated with the hole
when a spin pair is formed. Again, the holon having the effective Berry phase
factor $|\m_{eff}| = 1$ will also eventually form a pair of holes. This is
because of
the fact that as we see from eqn.(11), for any integer $\m$ the Berry phase
may be removed to the dynamical phase when the average magnetic field
vanishes. The Berry phase is observed when the system forms a pair
such that the units of magnetic flux are distributed among the pair.
The spinons $s_{i\sigma}$ minimizes the energy by acquiring a
pair amplitude given by
\beq
b_{ij}=s^{\dagger}_{i\uparrow }s^{\dagger}_{j\downarrow }-
s^{\dagger}_{i\downarrow }s^{\dagger}_{j\uparrow }
\eeq
Now we note that a spin pair each having unit magnetic flux and a pair of
holes each having unit magnetic flux should interact with each other through
the  gauge field $A_{ij}$ mentioned earlier.

It is now noted that the spin pairing as well as charge pairing in this
scheme occurs through a gauge interaction. In case of spin pairing, we note
that when the units of magnetic flux associated with the spinon having
$|\m_{eff}|=1$ are transferred to the hole, the residual spinon having
$|\m_{eff}| = 0$ eventually forms a pair of spins having $\m = 1/2$ and $-1/2$.
The magnetic flux associated with each spin will give rise to a gauge field
$A_{\m}$, lying in the link of the lattice, operating between them. Indeed,
the Berry phase factor is associated
with the
chiral anomaly through the relation [8]
\beq
2 \m =- \frac{ 1}{2} \int \pad_\m J^5_\m d^4x
\eeq
where $J^5_\m$ is the axial vector current $\bar{\psi} \g_\m \g_5 \psi$.
The association of a chiral current with spin is shown elsewhere [9]. When a
chiral current
interacts with a gauge field we have the anomaly given by
\beq
\pad_\m J^5_\m= Tr ~^\ast F_{\m\n} F_{\m\n}
\eeq
where $F_{\m \n}$ is  the field strength tensor of the gauge field $A_{\m}$.
This field can be associated with the background magnetic field. Thus we may
conclude that the gauge field $A_{ij}$ which is of magnetic type is
responsible for the spin-pairing observed in
high-$T_c$ superconductivity. The same view will also be valid for a pair of
holes which is eventually formed when the holon gets its share of magnetic flux
having $|\m_{eff}|=1$ from the spinon. This magnetic interaction is
responsible for the hole pairing which is strong enough to overcome the bare
Coulomb repulsion. This leads to the suggestion that the superconducting phase
order will be established when a spin pair each having unit magnetic flux and
a pair of holes having unit magnetic flux interacts with each other through a
gauge force. That is, the pair of holes will be attached to the spin pair such
that spin-charge recombination occurs when each hole
 is attached to a spin site of the spin pair.

 Also it is noted that the
bosonic holon having $|\m_{eff}| = 1$ and the residual {\it bosonic} spinon
having $|\m_{eff}| = 0$ (which eventually represents a pair)
cannot give the correct statistics for electron when these two form a
composite state. The correct statistics is only achieved when we introduce a
phase associated with a unit of magnetic flux corresponding to $\m = 1/2$ in
this composite system. Thus the spinon holon recombination along with a phase
shift only gives rise to an electron. This corroborates with the spin-charge
separation description in a path integral formalism [10] where an electron is
described as a composite particle of a spinon and holon together with  a
nonlocal phase-shift field.

From this approach, it appears that superconductivity and magnetism are
closely related. Indeed spinon-holon interaction
as well as the pair interaction is found to be of magnetic origin as the
magnetic flux associated with the Berry phase is responsible for these
features. Also we may infer about the phase diagram of the high
$T_c$ superconductiong states.
 As the number of holes are increased the number of
spinon-holon
recombined pair also increases, reaches a maximum at the optimally doped
region.
But as the percentage of the doped hole increases, in the overdoped region it
may not find the
spinon-pair to share its magnetic
flux and the spin-charge recombination stops. This may explain the downfall of
the phase diagram of high $T_c$ superconductors in the overdoped region.

{\bf Acknowledgement :} Authors are highly grateful to Prof. P. Bandyopadhayay
for helpful discussions. One of the authors D.P. is also grateful to CSIR for
financial support in doing this work.

\vspace{0.5cm}

\centerline{REFERENCES}

\vspace{1cm}

\begin{description}

\item{} [1] B. Basu, D. Pal and P. Bandyopadhyay : Int. J. Mod. Phys. B {\bf
13},(1999), 3393.
\item{} [2(a)] R. B. Laughlin : Science {\bf 242}, 525, (1988)
\item{} [2(b)]V. Kalmeyer and R. B. Laughlin : Phys. Rev. Lett. {\bf 59},
2095, (1987)
\item{} [3] P. Wiegmann : Prog. Theor. Phys. Suppl. {\bf 101}, 243, (1992)
\item{} [4] P. Bandyopadhyay : Int. J. Mod. Phys. A {\bf 15}, 1415, (2000)
\item{} [5] B. Basu and P. Bandyopadhyay : Int. J. Mod Phys. {B 11}, 2707,
(1997)
\item{} [6] B. Basu, D.Banerjee and P. Bandyopadhyay : Phys. Lett.
{\bf A, 236} (1997), 125
\item{} [7] Z. Y. Weng, D. N. Sheng and C. S. Ting : Phys. Rev. {\bf B 52},
637,(1995)
\item{} [8] B. Basu and P. Bandyopadhyay : Int. J. Mod Phys. {\bf B 12}, 2649,
(1998)
\item{} [9] G. Goswami and P. Bandyopadhyay : J. Math. Phys. {\bf 33}, 1090,
(1992).
\item{} [10] Z. Y. Weng, C. S. Ting and T. K. Lee : Phys. Rev B {\bf 43}, 3790
(1991)
\end{description}

\end{document}